\documentclass[a4paper,10pt,pdfusetitle]{article}
\usepackage[utf8]{inputenc}
\usepackage[T1]{fontenc}
\usepackage{lmodern}
\usepackage[svgnames]{xcolor}
\usepackage{amsmath,amssymb,amsfonts,mathrsfs,mathtools,graphicx}

\usepackage{titling}
\usepackage[margin=10pt,font=small,labelfont=bf,labelsep=colon]{caption}

\usepackage{geometry}
\usepackage{hyperref}
\hypersetup{colorlinks=true,allcolors=blue,linktocpage=true}
\usepackage{microtype}

\usepackage[numbers,sort&compress]{natbib}

\usepackage[color=lightgray]{todonotes}
\usepackage[normalem]{ulem}
\usepackage{authblk}

\usepackage{tikz}
\usetikzlibrary{positioning,fit}
\usetikzlibrary{decorations.pathmorphing}
\usetikzlibrary{decorations.markings}
\usetikzlibrary{calc}
\usetikzlibrary{arrows}

\tikzset{
  vertical align/.style={baseline={([yshift=-.5ex]current bounding box.center)}},
  scalar prop/.style={ultra thick, line cap=round},
  midarrow/.style={postaction={decorate}, decoration={
      markings,
      mark=at position 0.50 with {\arrow{>}}}},
  every picture/.style={vertical align, node distance=2.0cm}
}

\newcommand{\bb}[1]{\mathbb{#1}}
\newcommand{\id}{\mathrm{d}}

\title{Cuts and Isogenies}

\author[]{Hjalte~Frellesvig\thanks{hjalte.frellesvig@nbi.ku.dk}\thanksgap{1ex}}
\author[]{Cristian~Vergu\thanks{c.vergu@nbi.ku.dk}\thanksgap{1ex}}
\author[]{Matthias~Volk\thanks{mvolk@nbi.ku.dk}\thanksgap{1ex}}
\author[]{Matt~von~Hippel\thanks{mvonhippel@nbi.ku.dk}\thanksgap{1ex}}
\affil[]{{\small Niels Bohr International Academy \authorcr
Niels Bohr Institute \authorcr
University of Copenhagen \authorcr
Blegdamsvej 17, 2100 K{\o}benhavn, Denmark}}

\pretitle{\begin{center}\LARGE\bfseries}
\posttitle{\par\end{center}\vskip 2.5em}
\preauthor{%
\begin{center}
\large \lineskip 0.5em%
\begin{tabular}[t]{c}
}
\postauthor{\end{tabular}\par\end{center}\vskip 2.5em}
\predate{\begin{center}\large}
\postdate{\par\end{center}}

\thanksmarkseries{arabic}

\begin{document}

\begin{titlingpage}

\setlength{\droptitle}{0pt}

\maketitle

\begin{abstract}
    We consider the genus-one curves which arise in the cuts of the sunrise and in the elliptic double-box Feynman integrals.  We compute and compare invariants of these curves in a number of ways, including Feynman parametrization, lightcone and Baikov (in full and loop-by-loop variants).  We find that the same geometry for the genus-one curves arises in all cases, which lends support to the idea that there exists an invariant notion of genus-one geometry, independent on the way it is computed.  We further indicate how to interpret some previous results which found that these curves are related by isogenies instead.
\end{abstract}

\end{titlingpage}

\tableofcontents

\section{Introduction}
\label{sec:introduction}

There has recently been a flurry of interest in Feynman integrals associated with elliptic curves. Many different ways to represent these integrals have been developed~\cite{Broadhurst:1993mw,Laporta:2004rb,Bailey:2008ib,MullerStach:2011ru,Adams:2013nia,Bloch:2013tra,Broedel:2014vla,Adams:2015gva,Broedel:2015hia,Adams:2015ydq,Passarino:2016zcd,Primo:2016ebd,Remiddi:2016gno,Bonciani:2016qxi,Adams:2017ejb,vonManteuffel:2017hms,Ablinger:2017bjx,Remiddi:2017har,Hidding:2017jkk,Broedel:2017jdo,Lee:2017qql,Adams:2018bsn,Adams:2018kez,Ablinger:2018zwz}, culminating in bases of functions that are believed to be powerful enough to represent all such integrals~\cite{brown2011multiple,Broedel:2017kkb,Broedel:2018qkq}. A common feature of most of these representations is the characterization of each integral in terms of a single, specific elliptic curve. With the curve specified, relations can be found between functions defined on the same curve, allowing for the choice of a linearly independent basis.

What these representations typically do not consider are relations between Feynman integrals associated with different elliptic curves. This deficit is thrown into sharp relief by a pair of papers, one by Adams and Weinzierl~\cite{Adams:2017ejb}, and the other by Bogner, M\"uller-Stach, and Weinzierl~\cite{Bogner:2019lfa}, investigating the two-loop sunrise integral with all equal masses and with distinct internal masses respectively. These integrals have long been known to involve elliptic curves~\cite{Broadhurst:1993mw,Berends:1993ee,Bauberger:1994nk,Bauberger:1994by,Bauberger:1994hx,Caffo:1998du,Laporta:2004rb,Groote:2005ay,MullerStach:2011ru,Groote:2012pa,Adams:2013nia,Remiddi:2013joa,Bloch:2013tra,Adams:2014vja,Bloch:2014qca,Adams:2015gva,Bloch:2016izu,Broadhurst:2016myo,Remiddi:2016gno,Bogner:2017vim,Broedel:2017siw,Groote:2018rpb}. What they found was that the sunrise integral can in fact be described by two distinct elliptic curves in different contexts, with the curves related by a quadratic transformation, characterized in the latter paper as an isogeny~\cite{Bogner:2019lfa}. One curve appeared when analyzing the integral in terms of its Feynman-parametric representation, while another emerged from the maximal cut expressed in the Baikov representation~\cite{Baikov:1996iu} (see also~\cite{Lee:2010wea, Grozin:2011mt, Larsen:2015ped, Bosma:2017hrk, Frellesvig:2017aai, Harley:2017qut}). They refer to these as the curve from the graph polynomial and the curve from the maximal cut, respectively.

In this work, we investigate the origin of the distinction between these two curves: whether they differ because one comes from the maximal cut, or due to their origin in different representations. We examine two diagrams, the sunrise with all distinct internal masses and the elliptic double-box~\cite{CaronHuot:2012ab,Bourjaily:2017bsb}, in a variety of representations. In particular, we compare maximal cuts of these diagrams both in Baikov representations and in other representations (a light-cone representation in two dimensions, and a momentum twistor representation in four dimensions). We find that in general these representations can all give identical elliptic curves. Instead, we explain the observations of refs.~\cite{Adams:2017ejb,Bogner:2019lfa} as a consequence of a particular choice those references made when extracting an elliptic curve from the Baikov representation, involving combining two square roots. If we instead rationalize one of the roots, we find not an isogenous curve, but an \emph{identical} curve to that found in Feynman parametrization.

The paper is organized as follows: after a quick review of the relevant mathematics in section \ref{sec:review}, in section \ref{sec:sunrise} we consider the sunrise integral with three distinct masses. We review the Feynman-parametric representation in subsection \ref{sec:sunrise-feynman-parameters}, and the loop-by-loop Baikov representation found in ref.~\cite{Bogner:2019lfa} in subsection \ref{sec:sunrise-loop-by-loop}. We then derive two more representations, the traditional Baikov representation in subsection \ref{sec:sunrise-full-baikov} and a representation in light-cone coordinates in subsection \ref{sec:sunrise-lightcone}, and compare the resulting curves. In subsection \ref{sec:sunrise-rationalizing} we explain the differing curves as a result of combining distinct square roots, and extract an alternate curve by rationalizing a quadratic root instead, finding consistency with other methods. We give another view on the relation between the curves that avoids introducing square roots in subsection~\ref{sec:sunrise-double-cover}. In subsection \ref{sec:sunrise-landau} we close with a brief discussion of how the elliptic $j$-invariants of these curves shed light on the singularities of the diagram. In section \ref{sec:double-box} we investigate the elliptic double-box, where we compute Baikov representations of the maximal cut to compare to curves extracted in prior work. Specifically, we compare a \(d\)-dimensional Baikov representation (subsection~\ref{sec:double-box-baikov}) and a Baikov representation derived in strictly four dimensions (subsection~\ref{sec:4d-baikov}) finding agreement between the two. We then conclude and raise some topics for future investigation in section~\ref{sec:conclusions}.

Our paper also includes an appendix, reviewing both the loop-by-loop and the standard approach to the Baikov representation in~\ref{sec:baikov-derivations-loop-by-loop} and~\ref{sec:baikov-derivations-standard} respectively, as well as deriving our $d$-dimensional Baikov representation of the elliptic double-box in~\ref{sec:baikov-derivations-doublebox} and presenting more details of our four-dimensional derivation in~\ref{sec:4d-baikov-appendix}. We also include two ancillary files: \texttt{doublebox\_curve.txt}, presenting the elliptic curve for the double-box, and \texttt{doublebox\_baikov\_rep.txt}, presenting the Baikov representation for the double-box.

\section{Lightning review: Elliptic curves and isogenies}
\label{sec:review}

An elliptic curve is a smooth projective algebraic curve of genus one, together with a rational point which serves as the origin for its group structure.

There are many ways to represent such curves. One can write them as the vanishing loci of cubic polynomials in projective plane, or in terms of a quartic in a single variable with no repeated roots. One standard form is the so-called Weierstrass normal form, the equation
\begin{equation}
  \label{eq:weierstrass-model}
  y^2=4x^3-g_2 x-g_3\,,    
\end{equation}
for some coefficients $g_2$ and $g_3$.

Two elliptic curves are called isogenous when there is a non-constant map between them given by rational functions which sends the origin of the first to the origin of the second.  To every isogeny corresponds a dual isogeny and their composition is a homomorphism from an elliptic curve to itself.  If this homomorphism is the multiplication by two, we call the initial isogeny a two-isogeny.
If an isogeny has an inverse (that is, when the inverse map is also rational), one further calls the two curves isomorphic~\cite{lmfdb}. Isomorphic curves have the same $j$-invariant, which can be specified in terms of the coefficients of the Weierstrass normal form as follows\footnote{The factor of \(1728 = 2^6 \times 3^3\) is required for various number theoretic reasons which will not be relevant for us.  We choose to keep it in order to minimize confusion, but also because some of the formulas we will find below actually look \emph{nicer} when including this factor.}
\begin{align}
  j = 1728 \frac{g_2^3}{\Delta},
\end{align}
where the elliptic discriminant $\Delta=g_2^3-27g_3^2$.
The elliptic curve defined by the Weierstrass model~\eqref{eq:weierstrass-model} is smooth if and only if \(\Delta \neq 0\).

\section{The elliptic sunrise integral}
\label{sec:sunrise}

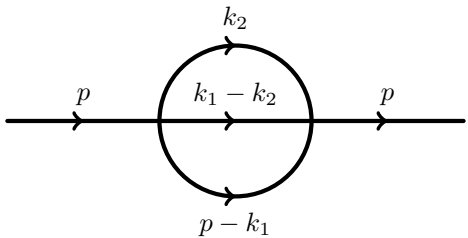
\begin{figure}[t]
  \centering
  \begin{tikzpicture}[]
    \node (X0) {};
    \node (X1) [right of=X0] {};
    \node (X2) [right of=X1] {};
    \node (X3) [right of=X2] {};
    \draw [scalar prop,midarrow] (X0.center) -- node[above=0.2em,midway] {$p$} (X1.center);
    \draw [scalar prop,midarrow] (X1.center) -- node[above=0.2em,midway] {$k_1 - k_2$} (X2.center);
    \draw [scalar prop,midarrow] (X2.center) -- node[above=0.2em,midway] {$p$} (X3.center);
    \draw [scalar prop,midarrow] (X1) arc (+180:+90:1) node[above=0.2em] {$k_2$} arc (+90:0:1);
    \draw [scalar prop,midarrow] (X1) arc (-180:-90:1) node[below=0.2em] {$p - k_1$} arc (-90:0:1);
  \end{tikzpicture}
  \caption{Sunrise integral. All internal propagators are massive and we consider the most general case where all masses can be unequal.  The momentum labeling is chosen such as to make the loop-by-loop Baikov representation easier to derive.}
  \label{fig:sunrise}
\end{figure}

The two-loop sunrise integral shown in fig.~\ref{fig:sunrise} is given by
\begin{align}
  \label{eq:sunrise-int-definition}
  I(p^2, m_1^2, m_2^2, m_3^2) =
  \int \frac{\mathrm{d}^2 k_1 \mathrm{d}^2 k_2}{\left(k_2^2 - m_1^2\right) \left((k_1 - k_2)^2 - m_2^2\right) \left((p - k_1)^2 - m_3^2\right)}.
\end{align}

This integral is finite in two dimensions, so it is often studied in that context.
In this section we will extract an elliptic curve from this integral in several ways, constructing the $j$-invariant for each such curve. We will find that the different methods we use provide only two distinct $j$-invariants, and are grouped as follows:

\begin{itemize}
\item Feynman parametrization (subsection~\ref{sec:sunrise-feynman-parameters}), solving the cut equations in light-cone coordinates (subsection~\ref{sec:sunrise-lightcone})
\item Loop-by-loop Baikov representation with 4 inverse propagators (subsection~\ref{sec:sunrise-loop-by-loop}), full Baikov representation with 5 inverse propagators (subsection~\ref{sec:sunrise-full-baikov})
\end{itemize}

These two $j$-invariants correspond to two distinct elliptic curves, which are not isomorphic.
However, as described in~\cite{Bogner:2019lfa}, the two curves are related by a two-isogeny.

In the rest of this section, we will describe how to extract an elliptic curve using each of these methods, and finish by reconciling the Baikov representations with the first set of methods, before briefly discussing this integral's Landau singularities.

\subsection{Feynman-parametric representation}
\label{sec:sunrise-feynman-parameters}

We begin by reviewing the two representations considered in ref.~\cite{Bogner:2019lfa}. The first representation considered in that reference was for the full integral expressed in Feynman parameters. In Feynman parameters, the integral can be written as \(\int \frac{\omega}{\mathfrak{F}}\) where \(\mathfrak{F}\) is the second graph polynomial,
\begin{align}
  \label{eq:fpoly-defining-eq}
  \mathfrak{F} = m_1^2 x_1^2 (x_2 + x_3) + m_2^2 x_2^2 (x_3 + x_1) + m_3^2 x_3^2 (x_1 + x_2)
      + (-p^2 + m_1^2 + m_2^2 + m_3^2) x_1 x_2 x_3
\end{align}
and
\begin{align}
  \omega = x_1 \id x_2 \id x_3 - x_2 \id x_1 \id x_3 + x_3 \id x_1 \id x_2.
\end{align}
The variables \(x_1\), \(x_2\) and \(x_3\) are homogeneous coordinates on \(\bb{P}^2\) and the equation \(\mathfrak{F} = 0\) defines an elliptic curve in \(\bb{P}^2\).\footnote{In this paper we always write \(\bb{P}^n\) for the complex projective space \(\bb{P}^n(\bb{C})\).}

To compute the $j$-invariant of this curve we may first divide by $p^2$ to make the expression dimensionless, then transform to the Weierstrass normal form.
For the purpose of writing the $j$-invariant for this curve, we define the following notation: writing \(\mu_i^2 = \frac{m_i^2}{p^2}\), we then write,
\begin{align}
  \label{eq:xis-definition}
  \xi_0 = \mu_1 + \mu_2 + \mu_3, \quad
  \xi_1 = -\mu_1 + \mu_2 + \mu_3, \quad
  \xi_2 = \mu_1 - \mu_2 + \mu_3, \quad
  \xi_3 = \mu_1 + \mu_2 - \mu_3.
\end{align}
With this notation, we can specify the $j$-invariant:
\begin{equation}
    j_F=\frac{\left[(\xi_0^2 - 1) (\xi_1^2 - 1) (\xi_2^2 - 1) (\xi_3^2 - 1) + 16 \mu_1^2 \mu_2^2 \mu_3^2\right]^3}{\mu_1^4 \mu_2^4 \mu_3^4 \, (\xi_0^2 - 1) (\xi_1^2 - 1) (\xi_2^2 - 1) (\xi_3^2 - 1)}
     \label{eq:paramj}
\end{equation}
where we have used a subscript \(F\) to indicate that this is computed from the Feynman parameter representation.

\subsection{Loop-by-loop Baikov representation}
\label{sec:sunrise-loop-by-loop}

Ref.~\cite{Bogner:2019lfa} presented the maximal cut of the two-loop sunrise integral in a loop-by-loop Baikov representation (as distinct from the traditional, or ``full'' Baikov representation, see ref.~\cite{Frellesvig:2017aai}, app.~\ref{sec:baikov-dervations-appendix}, or the next section to clarify the difference). We review below how to derive this representation in the case of this integral. 

In the Baikov representation we want to change the integration variables in the integral \(I(p^2, m_1^2, m_2^2, m_3^2)\) from the loop momenta \(k_1\) and \(k_2\) to the inverse propagators.
For the integral in eq.~\eqref{eq:sunrise-int-definition} the inverse propagators are
\begin{align}
  D_1 = k_2^2 - m_1^2, \quad
  D_2 = (k_1 - k_2)^2 - m_2^2, \quad
  D_3 = (p - k_1)^2 - m_3^2, \quad
  D_4 = k_1^2,
\end{align}
where we had to add \(D_4\) to be able to express all scalar products between the momenta.
In the following we consider the integral in the Euclidean region which corresponds to \(p^ 2 < 0\) and \(m_i^2 > 0\) for all masses.

The first step is to decompose the loop momenta into a part that is parallel and one that is orthogonal to the external momentum \(p\):
\begin{align}
  k_1 = x p + k_{1, \perp}, \quad
  k_2 = y p + k_{2, \perp}.
\end{align}
The orthogonal parts satisfy \(p \cdot k_{i, \perp} = 0\).
As we are in two dimensions, \(k_{1,\perp}\) and \(k_{2,\perp}\) are proportional and we can write them as \(k_{1, \perp} = u p_{\perp}\) and \(k_{2, \perp} = v p_{\perp}\).
Here \(p_{\perp}\) is chosen so that \(p \cdot p_{\perp} = 0\) and \(p_{\perp}^2 = p^2\).
Expressing the inverse propagators in terms of the dimensionless quantities \(x\), \(y\), \(u\) and \(v\) we obtain
\begin{align}
D_1 &= p^2 (y^2 + v^2) - m_1^2\,, \;
  & D_2 &= p^2 (x - y)^2 + p^2 (u - v)^2 - m_2^2\,, \nonumber \\
D_3 &= p^2 (x - 1)^2 + p^2 u^2 - m_3^2 \,, \; & D_4 &= p^2 (x^2 + u^2)\,.
\label{eq:sunrise-baikov-propagators-new-vars}
\end{align}
Moreover, the integration measure becomes \(\mathrm{d}^2 k_1 \mathrm{d}^2 k_2 = p^4 \, \mathrm{d}x \, \mathrm{d}y \, \mathrm{d}u \, \mathrm{d}v\).

We now want to change integration variables from \((x, y, u, v)\) to \((D_1, D_2, D_3, D_4)\) under which the measure transforms as \(\id x \, \id y \, \id u \, \id v = J^{-1} \, \id D_1 \, \id D_2 \, \id D_3 \, \id D_4\).
For the Jacobian factor \(J\) we get
\begin{align}
  \label{eq:sunrise-loop-by-loop-jacobian}
  J \equiv \left|\frac{\partial(D_1, D_2, D_3, D_4)}{\partial(x, y, u, v)}\right|
  = -16 p^8 u (u y - v x).
\end{align}
This Jacobian now has to be expressed in terms of the new variables \(D_i\).
The equations~\eqref{eq:sunrise-baikov-propagators-new-vars} are quadratic in \((x, y, u, v)\) and \(J\) can therefore not be expressed rationally in terms of the \(D_i\).
However, one can solve for the squares of \(u\) and \(u y - v x\) in eq.~\eqref{eq:sunrise-loop-by-loop-jacobian} rationally.
While this is possible for the full integral, here we only give the expression for the maximal cut corresponding to \(D_1 = D_2 = D_3 = 0\):\footnote{By abuse of notation we are here writing \(p\) for the absolute value of the momentum \(p^\mu\).}
\begin{align}
  \label{eq:sunrise-loop-by-loop-jfactors}
  \begin{split}
    Q_1 &:= u^2 = -\frac{1}{4 p^4} \left[D_4 - (m_3 - p)^2\right] \left[D_4 - (m_3 + p)^2\right], \\
    Q_2 &:= (u y - v x)^2 = -\frac{1}{4 p^4} \left[D_4 - (m_1 + m_2)^2\right] \left[D_4 - (m_1 - m_2)^2\right].
  \end{split}
\end{align}
Note that in the Euclidean region \(p^2\) is negative implying that the equation \(D_1 = 0\) does not have a real solution. In order to impose the cut conditions we are thus forced to consider the analytic continuation of the integral.

Multiplying \(Q_1\) and \(Q_2\) from the previous two equations we obtain an expression for \(J^2\) as a polynomial of degree four in \(D_4\).
This approach was followed in refs.~\cite{Adams:2017ejb,Bogner:2019lfa} and is equivalent to extracting the square root of each line in eq.~\eqref{eq:sunrise-loop-by-loop-jfactors} and combining the square roots under a common square root, i.e.\ to writing \(J = -16 p^8 \sqrt{Q_1 Q_2}\).
Another, inequivalent approach is to keep the square roots separate, i.e.\ to write \(J = -16 p^8 \sqrt{Q_1} \sqrt{Q_2}\).
As \(Q_1\) and \(Q_2\) are quadratic in \(D_4\), one can again change variables to rationalize either \(\sqrt{Q_1}\) or \(\sqrt{Q_2}\).
In subsection~\ref{sec:sunrise-rationalizing} we will show that this connects the elliptic curve arising from the first approach to the curve defined by the vanishing of the \(\mathfrak{F}\)-polynomial in subsection~\ref{sec:sunrise-feynman-parameters}.

Following the approach taken in ref.~\cite{Bogner:2019lfa}, we define an elliptic curve by the equation \(J^2 = (-16 p^8)^2 Q_1 Q_2\).
We can transform it to Weierstrass form and compute its $j$-invariant as in the previous section, obtaining:
\begin{equation}
    j_B=\frac{\left[(\xi_0^2 - 1) (\xi_1^2 - 1) (\xi_2^2 - 1) (\xi_3^2 - 1) + 256 \mu_1^2 \mu_2^2 \mu_3^2\right]^3}{\mu_1^2 \mu_2^2 \mu_3^2 \, (\xi_0^2 - 1)^2 (\xi_1^2 - 1)^2 (\xi_2^2 - 1)^2 (\xi_3^2 - 1)^2},
    \label{eq:baikovj}
\end{equation}
where we have again made use of \(\mu_i^2 = \frac{m_i^2}{p^2}\) and the variables \(\xi_i\) defined in eq.~\eqref{eq:xis-definition}.
This clearly differs from the $j$-invariant computed in the previous subsection, see eq.~\eqref{eq:paramj}. However, as observed in ref.~\cite{Bogner:2019lfa}, the two curves are isogenous.  This has been checked in ref.~\cite{Bogner:2019lfa} by computing the complex structure parameter \(\tau\) of the elliptic curve.  Here we check it by using the relations between the \(j\)-invariants of the two elliptic curves.  The \(j\)-invariants for a pair of two-isogenous elliptic curves are related by the modular polynomial \(\Phi_2(X, Y)\) (see e.g.~\cite[Chapter 5]{LangEllipticFunctions})
\begin{align}
  \begin{split}
    \Phi_2(X, Y)
    &= X^3 + Y^3 - X^2 Y^2 + 1488 \left(X^2 Y + X Y^2\right) - 162000 \left(X^2 + Y^2\right) \\
    &\qquad + 40773375 \, X Y + 8748000000 \left(X + Y\right) - 157464000000000.
  \end{split}
\end{align}
See ref.~\cite{MR2869057} for details about how these modular polynomials are computed.
It can be checked that \(\Phi_2(j_F, j_B) = 0\).  This is an infinite precision test of two-isogeny.  Ref.~\cite{Bogner:2019lfa} used the approach of comparing the periods which are computed using elliptic integrals.  This involves transcendental functions while the approach we followed here only requires algebraic operations with rational functions.

\subsection{Full Baikov representation}
\label{sec:sunrise-full-baikov}

For a ``full'' Baikov approach to an \(L\)-loop integral with \(E + 1\) external legs one needs \(\frac{1}{2} L (L + 1) + L E\) Baikov variables \(D_a\).
In the present case (\(L = 2\), \(E = 1\), \(M = L + E = 3\)), the variables are \(D_1, \ldots, D_5\) and the maximal cut corresponds to setting \(D_1 = D_2 = D_3 = 0\) at the end of the computation.

We now follow~\cite{Frellesvig:2017aai} to derive the Baikov representation.
The inverse propagators are
\begin{align}
\quad D_1 &= k_2^2 - m_1^2\;, & D_2 &= (k_1-k_2)^2 - m_2^2\;, & D_3 &= (p - k_1)^2 - m_3^2\;, \quad \nonumber \\
\quad D_4 &= k_1^2\;, & D_5 &= (p - k_2)^2\;. &
\end{align}
Loosely following the notation of the paper above we set \(q_1 = k_1\), \(q_2 = k_2\) and \(q_3 = p\) and write  \(s_{ij} = q_i \cdot q_j\).
The Gram determinant\footnote{The astute reader may notice that this Gram determinant vanishes when in strictly two dimensions. If one is uncomfortable with this one can instead derive a Baikov representation strictly in two dimensions. We will do something similar for the elliptic double-box in section~\ref{sec:4d-baikov}. Details relevant for either case (in particular, how to handle cases when the internal momenta are spanned by the external momenta) are presented in appendix~\ref{sec:4d-baikov-appendix}.} is
\begin{align}
  \begin{split}
    G(k_1, k_2, p) &=
    \det
    \begin{pmatrix}
      s_{11} & s_{12} & s_{13} \\
      s_{12} & s_{22} & s_{23} \\
      s_{13} & s_{23} & p^2
    \end{pmatrix} \\
    &= s_{11} \left(p^2 s_{22} - s_{23}^2\right) - s_{12} \left(p^2 s_{12} - s_{13} s_{23}\right) + s_{13} \left(s_{12} s_{23} - s_{13} s_{22}\right).
  \end{split}
\end{align}
The Baikov polynomial is obtained by rewriting the Mandelstam variables \(s_{ij}\) in terms of the inverse propagators \(D_a\) in this Gram determinant,
\begin{align}
  P(D_1, \ldots, D_5) = G(k_1, k_2, p) \Bigr|_{s_{ij}(D_a)}.
  \end{align}

The cut integral (\(D_1 = D_2 = D_3 = 0\)) is of the form
\begin{align}
  \label{eq:sunrise-full-baikov-cut}
  \int \frac{\mathrm{d} D_4 \mathrm{d} D_5}{D_4^{\alpha_4} D_5^{\alpha_5}} P(0, 0, 0, D_4 , D_5 )^{(d - M - 1) / 2}.
\end{align}
Where $\alpha_4$ and $\alpha_5$ are the exponents of $D_4$ and $D_5$ in the original integral respectively. Since \(M = 3\), \(d = 2\) and \(\alpha_4 = \alpha_5 = 0\) we get
\begin{align}
  \int \frac{\mathrm{d} D_4 \mathrm{d} D_5}{P(0,0,0,D_4, D_5)},
\end{align}
where \(P\) is a polynomial of overall degree three in \(D_4\) and \(D_5\),
\begin{align}
    P=\frac{1}{4}\Big[-D_4^2 D_5&+D_5(m_1^2-m_2^2)(m_3^2-p^2)-(m_1^2 m_3^2-m_2^2 p^2)(m_1^2-m_2^2+m_3^2-p^2)\nonumber\\
    &-D_4(D_5^2+(m_2^2-m_3^2)(m_1^2-p^2)-D_5(m_1^2+m_2^2+m_3^2+p^2))\Big].
\end{align}

The equation \(P = 0\) defines an elliptic curve. We may again transform this curve to Weierstrass form. As it turns out, this curve has the same \(j\)-invariant as that from the loop-by-loop Baikov computation in the previous section. Rather than repeating it here we thus refer back to eq.~(\ref{eq:baikovj}).

\subsection{Light-cone coordinates}
\label{sec:sunrise-lightcone}

One convenient way to enforce on-shell conditions in two dimensions is via light-cone coordinates. We wish to enforce the conditions for the maximal cut:
\begin{align}
  \label{eq:light-cone-cut-conditions}
  k_2^2 - m_1^2 = 0, \quad
  (k_1 - k_2)^2 - m_2^2 = 0, \quad
  (p - k_1)^2 - m_3^2 = 0.
\end{align}

We define the auxiliary momentum \(k_3 = k_1 - k_2\) and use that in light-cone coordinates the square of a momentum is given by \(k_i^2 = k_i^+ k_i^-\).
Then the first two conditions in eq.~\eqref{eq:light-cone-cut-conditions} are solved by
\begin{align}
  k_2^- = \frac{m_1^2}{k_2^+}, \quad
  k_3^- = \frac{m_2^2}{k_3^+}.
\end{align}
The last condition in eq.~\eqref{eq:light-cone-cut-conditions} becomes
\begin{align}
  (p^+ - k_2^+ - k_3^+) (p^- - k_2^- - k_3^-) - m_3^2 =
  (p^+ - k_2^+ - k_3^+) \left(p^- - \frac{m_1^2}{k_2^+} - \frac{m_2^2}{k_3^+}\right) - m_3^2 =
  0.
\end{align}
Introducing dimensionless quantities as \(k_2^+ = p^+ x\), \(k_3^+ = p^+ y\) and again using \(\mu_i^2 = \frac{m_i^2}{p^2}\), the previous equation becomes
\begin{align}
  (1 - x - y) \left(1 - \frac{\mu_1^2}{x} - \frac{\mu_2^2}{y}\right) - \mu_3^2
  = 0.
\end{align}
In homogeneous coordinates \([x : y : z]\) and after multiplying by \(x y z\) we are left with a cubic curve in \(\bb{P}^2\) given by the equation
\begin{align}
  \label{eq:defining-eq-lightcone}
  P_L \equiv
  x y z \left(1 + \mu_1^2 + \mu_2^2 - \mu_3^2\right)
  + x^2 \left(\mu_2^2 z - y\right)
  + y^2 \left(\mu_1^2 z - x\right)
  - z^2 \left(\mu_2^2 x + \mu_1^2 y\right)
  = 0.
\end{align}

This is an elliptic curve whose defining equation is closely related to the \(\mathfrak{F}\)-polynomial in~\eqref{eq:fpoly-defining-eq}.
Specifically, their discriminants with respect to \(z\) are related by
\begin{align}
  \label{eq:light-cone-feynman-relation}
  \operatorname{disc}_{z} P_L(x, y, z) = \operatorname{disc}_{z} \mathfrak{F}(y, x, z).
\end{align}
Once again we can transform the curve to Weierstrass form, and evaluate its $j$-invariant. As suggested by the relationship in eq.~\eqref{eq:light-cone-feynman-relation}, we find it has the same $j$-invariant as the Feynman parametric representation (given in eq.~\eqref{eq:paramj}), and a distinct (but isogenous) $j$-invariant to those in the two Baikov representations.

\subsection{Rationalizing the square roots in the Baikov representation}
\label{sec:sunrise-rationalizing}

In subsection~\ref{sec:sunrise-loop-by-loop} we derived a loop-by-loop Baikov representation of the sunrise integral and explained how the equation \(J = -16 p^8 \sqrt{Q_1 Q_2}\) defines an elliptic curve isogenous to the one obtained by Feynman parameters and the light-cone computation as in ref.~\cite{Bogner:2019lfa}.
Combining \(\sqrt{Q_1}\) and \(\sqrt{Q_2}\) in this way is safe if both \(Q_1\) and \(Q_2\) are positive.
However, for complex kinematics it may lead to an incorrect phase.

Instead of combining the two roots, we can rationalize one of them.
Recall that \(Q_1\) and \(Q_2\) were given in eq.~\eqref{eq:sunrise-loop-by-loop-jfactors} as
\begin{align}
  \begin{split}
    Q_1 &= -\frac{1}{4 p^4} \left[D_4 - (m_3 - p)^2\right] \left[D_4 - (m_3 + p)^2\right], \\
    Q_2 &= -\frac{1}{4 p^4} \left[D_4 - (m_1 + m_2)^2\right] \left[D_4 - (m_1 - m_2)^2\right].
  \end{split}
\end{align}
Choosing to rationalize \(\sqrt{Q_2}\), the change of variables amounts to replacing
\begin{align}
  \begin{split}
    D_4 &\to 2 t \left[\frac{m_1^2}{t - 1} + \frac{m_2^2}{t + 1}\right], \\
    \sqrt{Q_2} &\to \frac{(t (m_1 - m_2) + (m_1 + m_2)) (t (m_1 + m_2) + (m_1 - m_2))}{2 p^2 (t^2 - 1)}.
  \end{split}
\end{align}

It turns out that the Jacobian from the change of variables cancels against the transformed \(\sqrt{Q_2}\) and a factor of \(t^2 - 1\) coming from \(\sqrt{Q_1}\).
In the end we obtain
\begin{align}
  I(p^2, m_1^2, m_2^2, m_3^2)\Big|_{\mathrm{cut}}
  =
  p^4 \int \frac{\id D_4}{J}
  =
  -\frac{1}{16 p^4}
  \int \frac{\mathrm{d} D_4}{\sqrt{Q_1} \sqrt{Q_2}}
  =
  -\frac{1}{16 p^2}
  \int \frac{\mathrm{d} t}{\sqrt{R}},
\end{align}
where \(R\) is a polynomial of degree four in \(t\),
\begin{align}
  \begin{split}
    R &\equiv
    \frac{1}{64 p^4} \left[\left((m_3 - p)^2 - 2 (m_1^2 + m_2^2)\right) t^2 - 2 (m_1^2 - m_2^2) t - (m_3 - p)^2\right] \\
    &\hphantom{\equiv \frac{1}{64 p^4}\,} \times
    \left[\left((m_3 + p)^2 - 2 (m_1^2 + m_2^2)\right) t^2 - 2 (m_1^2 - m_2^2) t - (m_3 + p)^2\right].
  \end{split}
\end{align}
The equation \(y^2 = R(t)\) defines an elliptic curve as a hypersurface in a weighted projective space \(\bb{P}^{1:1:2}\).
It turns out that this curve has the same \(j\)-invariant as that from the graph polynomial (given in eq.~\eqref{eq:paramj}).
Note that this is \emph{not} the same \(j\)-invariant as in the Baikov representations above, even though the loop-by-loop Baikov representation was our starting point: by rationalizing instead of combining roots we have achieved agreement with the graph polynomial and light-cone derivations of the elliptic curve.

Another way to think about how the two curves emerge is to track what happens to the branch points of the curves under the change of variables above. In ref.~\cite{Bogner:2019lfa} and subsection~\ref{sec:sunrise-loop-by-loop} the elliptic curve arising from the Baikov representation is defined by \(J^2 = (-16 p^8)^2 Q_1 Q_2\).
This is a double cover of \(\bb{P}^1\) branched over four points.
Since \(Q_1\) and \(Q_2\) are already factorized, the branch points are easy to read off:
\begin{align}
    D_{4,\pm}^{(1)} = (m_1 \pm m_2)^2, \quad
    D_{4,\pm}^{(2)} = (m_3 \pm p)^2.
\end{align}
These are four points on a projective line parametrized by the coordinate \(D_4\).  They have a cross-ratio \(\lambda\) with corresponding \(j\)-invariant \(j = 256 \frac{(\lambda^2 - \lambda + 1)^3}{\lambda^2 (1 - \lambda)^2}\).
This approach gives the ``Baikov'' \(j\)-invariant shown in eq.~\eqref{eq:baikovj}.

On the other hand, when rationalizing the quadric \(Q_2\) we write \(x\) as the image of a map from a different \(\bb{P}^1\) with coordinate \(t\),
\begin{align}
    t \mapsto x(t) = 2 t \left[\frac{m_1^2}{t - 1} + \frac{m_2^2}{t + 1}\right].
\end{align}
Under this change of variables, \(Q_1\) becomes a polynomial of degree four and we again define an elliptic curve as a double cover of \(\bb{P}^1\), but this time the \(\bb{P}^1\) has coordinate \(t\).
The branch points are the preimages of the two points \(D_{4,+}^{(2)}\) and \(D_{4,-}^{(2)}\).
Since the change of variables is quadratic each point has two preimages and we indeed get four branch points as required.
As we now again have four points on a projective line, we can form a cross-ratio and the corresponding \(j\)-invariant.
This is the \(j\)-invariant that comes from the \(\mathfrak{F}\)-polynomial and the light-cone approach in eq.~\eqref{eq:paramj}.

The analysis presented here applies to the loop-by-loop Baikov representation, and at first this may make the full Baikov result seem mysterious, as unlike the loop-by-loop representation it does not obviously involve combining square roots. However, if one derives the Baikov representation by dividing each loop momentum into perpendicular and parallel subspaces, as for example in ref.~\cite{Grozin:2011mt}, then one naturally passes through a form closely related to the loop-by-loop representation in which there are indeed multiple square roots. In particular, the individual equations that need to be solved to land on the cut solution will be the same. If one understands the Baikov representation as a result of this kind of procedure, then the elliptic curve we found for it earlier can be explained in the same way as the loop-by-loop curve, and a similarly more careful treatment (especially one along the lines of the next section) will result in the same curve as was found from Feynman parameters and light-cone coordinates.

\subsection{Derivation of the double cover relation}
\label{sec:sunrise-double-cover}

In this section we study the relation between the two genus-one curves from a different point of view.  We describe the curves purely by polynomial equations and we avoid introducing square roots.

On the maximal cut we have \(D_1 = D_2 = D_3 = 0\) and these equations together with \(D_4 = p^2 (x^2 + u^2)\) define a curve.
We introduce a dimensionless variable \(d_4 = \frac{D_4}{p^2}\).
Then, the equations~\eqref{eq:sunrise-baikov-propagators-new-vars} can be simplified by solving
\begin{gather}
    x = \frac{d_4 - \mu_3^2 + 1}{2}, \\
    u^2 = -\frac 1 4 (d_4 - (1 + \mu_3)^2) (d_4 - (1 - \mu_3)^2), \\
    v^2 = \mu_1^2 - y^2, \\
    u v = (1 - y) \frac{d_4 - \mu_3^2 + 1}{2} + \frac{\mu_1^2 - \mu_2^2 + \mu_3^2 - 1}{2}.
\end{gather}
We now obtain the equation for the curve in variables \(y\) and \(d_4\), by substituting the expressions above in \((u v)^2 = u^2 v^2\).  This equation is
\begin{multline}
    P(y, d_4) = -4 y^{2} d_{4} + 2 y d_{4}^{2} + 2 \left(\mu_{1}^{2} - \mu_{2}^{2} - \mu_{3}^{2} + 1\right) y d_{4} - \left(\mu_{1}^{2} + 1\right) d_{4}^{2} +\\ + 2 \left(-\mu_{1}^{2} \mu_{3}^{2} + \mu_{2}^{2} \mu_{3}^{2} + \mu_{1}^{2} - \mu_{2}^{2}\right) y + 2 \left(\mu_{1}^{2} \mu_{3}^{2} + \mu_{2}^{2}\right) d_{4} \\ - \mu_{1}^{2} \mu_{3}^{4} - \mu_{1}^{4} + 2 \mu_{1}^{2} \mu_{2}^{2} - \mu_{2}^{4} + 2 \mu_{1}^{2} \mu_{3}^{2} - \mu_{1}^{2} = 0,
\end{multline}
and is a cubic equation in \(y\) and \(d_4\).  It is not in Weierstrass form.
The expression for the Jacobian can also be written in the variables \(y\) and \(d_4\):
\begin{equation}
    J = \frac{p^8}4 \bigl(4 y d_4 - (d_4 + \mu_1^2 - \mu_2^2) (d_4 - \mu_3^2 + 1)\bigr).
\end{equation}

Note that this approach avoids introducing square roots, at the cost of working with two variables constrained by an algebraic relation.

Let us show that \(\frac{\id d_4}{J}\) is the holomorphic one-form on this curve.  Taking the differential of \(P(y, d_4) = 0\) we obtain
\begin{equation}
    \left(\frac{\partial P}{\partial y}\right) \id y + \left(\frac{\partial P}{\partial d_4}\right) \id d_4 =
    -2 J(y, d_4) \, \id y + K(y, d_4) \, \id d_4 = 0,
\end{equation}
where
\begin{equation}
    K(y, d_4) = -4 y^{2} + 4 y d_{4} + 2 (\mu_{1}^{2} - \mu_{2}^{2} - \mu_{3}^{2} + 1) y - 2 ( \mu_{1}^{2} +1) d_{4} + 2 \mu_{1}^{2} \mu_{3}^{2} + 2 \mu_{2}^{2}.
\end{equation}

Since we assume that the curve described by \(P = 0\) is nonsingular, we have that \(\frac{\partial P}{\partial y} = -2 J\) and \(\frac{\partial P}{\partial d_4} = K\) can not vanish simultaneously.  Then, we have \(\frac{\id d_4}{J} = 2 \frac{\id y}{K}\).  Hence, one can see that at the zeros of \(J\) this holomorphic form does not have poles, when written with the denominator \(K\).  It can be checked that this curve is the same as the one obtained by the more traditional Baikov approach.

However, one can see that the curve we started with, in the variables \(x\), \(y\), \(u\), \(v\) and \(d_4\) is a double cover of the curve \(P(d_4, y) = 0\).  Given a point \((d_4, y)\), we can uniquely find \(x\) and \(u^2\), \(v^2\) and \(u v\).  This allows us to solve for \(u\) and \(v\) up to a sign.  Hence, to a point on the curve \(P(d_4, y) = 0\) correspond two points on the initial curve defined by \(D_1 = D_2 = D_3 = 0\) and \(d_4 = x^2 + u^2\).

To find a one-to-one projection of the curve which is easily recognizable as an elliptic curve, we proceed as follows.  We can use a kind of Euclidean lightcone construction and transform the equations to
\begin{gather}
    y + i v = \frac{\mu_1^2}{y - i v}, \\
    (x - y) + i (u - v) = \frac{\mu_2^2}{(x - y) - i (u - v)}, \\
    (x - 1) + i u = \frac{\mu_3^2}{(x - 1) - i u}.
\end{gather}
Combining them, we find
\begin{equation}
    \frac{\mu_1^2}{y - i v} + \frac{\mu_2^2}{(x - y) - i (u - v)} - \frac{\mu_3^2}{(x - 1) - i u} = 1.
\end{equation}
If we introduce \(\zeta = y - i v\) and \(\xi = x - i u\), we have a curve
\begin{equation}
    \frac{\mu_1^2}{\zeta} + \frac{\mu_2^2}{\xi - \zeta} - \frac{\mu_3^2}{\xi - 1} = 1,
\end{equation}
which is a cubic equation in \((\zeta, \xi)\).  Once we have \(\zeta\) and \(\xi\) we obtain
\begin{gather}
    y = \frac 1 2 \left(\zeta + \frac{\mu_1^2}{\zeta}\right), \qquad
    v = \frac 1 {2 i} \left(-\zeta + \frac{\mu_1^2}{\zeta}\right),
\end{gather}
and similarly for \(x\) and \(u\).  Finally, we obtain \(d_4 = x^2 + u^2\).  This time, given a point \((\zeta, \xi)\) we can find a unique point on the initial curve.

This second curve looks very similar to the lightcone solution of sec.~\ref{sec:sunrise-lightcone} and indeed it has the same \(j\)-invariant.

\subsection{Singularities of the geometry and Landau analysis}
\label{sec:sunrise-landau}

Recall that the \(j\)-invariant of an elliptic curve is
\begin{align}
  j = 1728 \frac{g_2^3}{\Delta},
\end{align}
where \(\Delta\) is the elliptic discriminant.
When \(\Delta\) vanishes, $j$ is singular, and the elliptic curve degenerates.

For the curve arising from Feynman parametrization and the light-cone computation we obtained
\begin{equation}
    j_F=\frac{\left[(\xi_0^2 - 1) (\xi_1^2 - 1) (\xi_2^2 - 1) (\xi_3^2 - 1) + 16 \mu_1^2 \mu_2^2 \mu_3^2\right]^3}{\mu_1^4 \mu_2^4 \mu_3^4 \, (\xi_0^2 - 1) (\xi_1^2 - 1) (\xi_2^2 - 1) (\xi_3^2 - 1)}
\end{equation}
while for the curve arising from the Baikov representation we obtained
\begin{equation}
    j_B=\frac{\left[(\xi_0^2 - 1) (\xi_1^2 - 1) (\xi_2^2 - 1) (\xi_3^2 - 1) + 256 \mu_1^2 \mu_2^2 \mu_3^2\right]^3}{\mu_1^2 \mu_2^2 \mu_3^2 \, (\xi_0^2 - 1)^2 (\xi_1^2 - 1)^2 (\xi_2^2 - 1)^2 (\xi_3^2 - 1)^2}.
\end{equation}

The denominators of these expressions are distinct, but they clearly have the same zeros, just with different multiplicities.
These zeros all correspond to physical singularities of the diagram, either to thresholds, pseudo-thresholds, or vanishing internal masses. Each corresponds to a consistent Landau diagram, for particular choices of the sign of the energies of each particle. The easiest to recognize are the thresholds, occurring when $(\xi_0)^2=1$ and thus $(m_1+m_2+m_3)^2=p^2$, which is the condition for energy conservation when all of the intermediate particles are traveling in the same direction. The Landau analysis also reveals that there are other singularities, arising at pseudo-thresholds \(p^2 = (-m_1 + m_2 + m_3)^2\), \(p^2 = (m_1 - m_2 + m_3)^2\) and \(p^2 = (m_1 + m_2 - m_3)^2\).  In terms of variables \(\xi\) these are \((\xi_1)^2 = 1\), \((\xi_2)^2 = 1\) and \((\xi_3)^2 = 1\). Finally, the singularities arising when one of the masses vanishes are of a different type.  They arise due to the fact that when one of the masses vanishes the integral becomes divergent.

\section{The elliptic double-box integral}
\label{sec:double-box}

The elliptic double-box integral has previously been analyzed in ref.~\cite{Bourjaily:2017bsb} from the point of view of direct integration in a Feynman parametric representation, and in ref.~\cite{Vergu:2020uur} from the point of view of the maximal cut in twistor space.
In both papers the same elliptic curve was found using very different methods. In this section we derive a Baikov representation of the double-box, and show that it also defines the same curve.

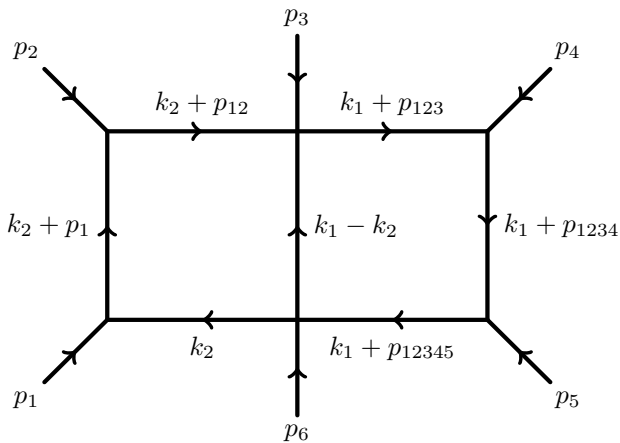
\begin{figure}[ht]
  \centering
  \begin{tikzpicture}[node distance=2.5cm]
    \node (A) {};
    \node (B) [right of=A] {};
    \node (C) [right of=B] {};
    \node (D) [below of=A] {};
    \node (E) [right of=D] {};
    \node (F) [right of=E] {};
    \node (Aext) at ($(A) + (+135:1.5)$) {$p_2$};
    \node (Bext) at ($(B) + ( +90:1.5)$) {$p_3$};
    \node (Cext) at ($(C) + ( +45:1.5)$) {$p_4$};
    \node (Dext) at ($(D) + (-135:1.5)$) {$p_1$};
    \node (Eext) at ($(E) + ( -90:1.5)$) {$p_6$};
    \node (Fext) at ($(F) + ( -45:1.5)$) {$p_5$};
    \draw [scalar prop,midarrow] (A.center) -- node[above=0.2em,midway] {$k_2 + p_{12}$} (B.center);
    \draw [scalar prop,midarrow] (B.center) -- node[above=0.2em,midway] {$k_1 + p_{123}$} (C.center);
    \draw [scalar prop,midarrow] (F.center) -- node[below=0.2em,midway] {$k_1 + p_{12345}$} (E.center);
    \draw [scalar prop,midarrow] (E.center) -- node[below=0.2em,midway] {$k_2$} (D.center);
    \draw [scalar prop,midarrow] (D.center) -- node[left=0.2em,midway] {$k_2 + p_1$} (A.center);
    \draw [scalar prop,midarrow] (E.center) -- node[right=0.2em,midway] {$k_1 - k_2$} (B.center);
    \draw [scalar prop,midarrow] (C.center) -- node[right=0.2em,midway] {$k_1 + p_{1234}$} (F.center);
    \draw [scalar prop,midarrow] (Aext) -- node[right,midway] {} (A.center);
    \draw [scalar prop,midarrow] (Bext) -- node[right,midway] {} (B.center);
    \draw [scalar prop,midarrow] (Cext) -- node[right,midway] {} (C.center);
    \draw [scalar prop,midarrow] (Dext) -- node[right,midway] {} (D.center);
    \draw [scalar prop,midarrow] (Eext) -- node[right,midway] {} (E.center);
    \draw [scalar prop,midarrow] (Fext) -- node[right,midway] {} (F.center);
  \end{tikzpicture}
  \caption{Double-box integral in momentum space. Incoming momenta are assumed to be off-shell, i.e. \(p_i^2 \neq 0\), and $p_{i_1 \cdots i_n} \equiv p_{i_1} + \cdots + p_{i_n}$. The internal propagators are massless.}
  \label{fig:double-box}
\end{figure}

\subsection{Baikov representation}
\label{sec:double-box-baikov}

The Baikov representation is a rewriting of Feynman integrals where the integration is over Lorentz-invariant quantities, such as dot products.
In appendix~\ref{sec:baikov-dervations-appendix} we derive such a representation for the elliptic double-box integral shown in fig.~\ref{fig:double-box} (see in particular appendix~\ref{sec:baikov-derivations-doublebox}).

The maximal cut of the elliptic double-box can be written in a loop-by-loop Baikov representation as an integral over two Baikov parameters. The cut integrand takes the following form:
\begin{align}
  \label{eq:doublebox-baikov-integrand}
  \frac{\mathcal{J} \sqrt{\mathcal{G}_1} \, \id x_8 \id x_9}{\mathcal{B}_1(x_8, x_9) \sqrt{\mathcal{B}_2(x_8, x_9)}},
\end{align}
where \(x_8\) and \(x_9\) are the two remaining Baikov variables after all propagators have been cut.
The polynomials \(\mathcal{B}_1\) and \(\mathcal{B}_2\) are of degree two in \(x_8\) and also of degree two in \(x_9\).
The factors \(\mathcal{J}\) and \(\mathcal{G}_1\) only depend on the external kinematics. We include expressions for these polynomials in an ancillary file, \texttt{doublebox\_baikov\_rep.txt}.

To obtain an elliptic curve, we may begin by taking a residue around \(\mathcal{B}_1(x_8, x_9) = 0\). Without loss of generality, let us take this residue in \(x_9\).
Solving \(\mathcal{B}_1(x_8, x_9) = 0\) for \(x_9\) introduces a square root that contains \(x_8\), and this square root can be rationalized by Euler substitution as done in subsection~\ref{sec:sunrise-rationalizing} for the sunrise integral.
Denoting by \(t\) the variable that replaces \(x_8\) to rationalize the square root we find that \(\mathcal{B}_2(t)\) becomes a quartic polynomial in \(t\).
We can therefore define an elliptic curve by \(y^2 = \mathcal{B}_2(t)\) and compute its \(j\)-invariant through standard changes of variables.

The problem with this approach is that the change of variables from \(x_8\) to \(t\) may itself introduce a square root in the kinematic parameters.
Since the \(j\)-invariant of the elliptic curve is expected to be a rational function of the kinematics, this root is spurious and must cancel in \(j\).

The spurious kinematic root can be avoided if we view \(x_8\) and \(x_9\) as a subset of the coordinates on a \(\bb{P}^3\) with homogeneous coordinates \([x_8 : x_9 : y : z]\).
From the denominator in the integrand in eq.~\eqref{eq:doublebox-baikov-integrand} we define the two quadrics\footnote{Note that here we are writing \(\mathcal{B}_1(x_8, x_9, z)\) for the homogenization of the polynomial \(\mathcal{B}_1\) in eq.~\eqref{eq:doublebox-baikov-integrand} and similarly for \(\mathcal{B}_2\).}
\begin{align}
  \label{eq:doublebox-baikov-quadrics}
  \begin{split}
  Q_1&: \{[x_8 : x_9 : y : z] \in \bb{P}^3 \mid \mathcal{B}_1(x_8, x_9, z) = 0\}, \\
  Q_2&: \{[x_8 : x_9 : y : z] \in \bb{P}^3 \mid y^2 - \mathcal{B}_2(x_8, x_9, z) = 0\}.
  \end{split}
\end{align}
The integrand in eq.~\eqref{eq:doublebox-baikov-integrand} corresponds to a differential form on the intersection of \(Q_1\) and \(Q_2\).
For generic quadrics \(Q_1\) and \(Q_2\) this intersection is a smooth curve of genus one.

We now review briefly how this curve may be characterized and refer to~\cite[Chapter 22]{MR1182558} for further details.
The quadrics \(Q_1\) and \(Q_2\) generate a family of quadrics
\begin{align}
  \label{eq:pencil-of-quadrics}
  \{\lambda_0 Q_1 + \lambda_1 Q_2 \mid [\lambda_0: \lambda_1] \in \bb{P}^1\}.
\end{align}
This family is called the \emph{pencil of quadrics} and the intersection \(C = Q_1 \cap Q_2\) is called the \emph{base locus} of the pencil.
The members \(Q_\lambda\) of the pencil are quadrics in \(\bb{P}^3\) and for some choices of \(\lambda \in \bb{P}^1\) they may be singular.
If \(Q_1\) and \(Q_2\) intersect transversely, there are four such singular members \(Q_{\lambda_i}\) with \(i = 0, \ldots, 3\).
Out of the four points \(\lambda_i\) we can form a cross-ratio \(\kappa\) and subsequently the invariant combination
\begin{align}
  j = 256 \frac{(\kappa^2 - \kappa + 1)^3}{\kappa^2 (\kappa - 1)^2},
\end{align}
which characterizes the pencil of quadrics up to isomorphism.
One can now moreover show that the base locus \(C\) of the pencil is isomorphic to a genus-one curve in the plane with the same \(j\)-invariant as the pencil.

An advantage of this description is that it allows us to compute the elliptic discriminant of the curve using only rational operations.
Writing \(Q_1\) and \(Q_2\) for the \(4 \times 4\) symmetric matrices associated to the quadrics in eq.~\eqref{eq:doublebox-baikov-quadrics} the locations \(\lambda_i\) of the singular members of the pencil are given by the eigenvalues of the matrix \(Q_2^{-1} Q_1\).
The curve degenerates if two of those points in \(\bb{P}^1\) are the same, i.e.\ if \(Q_2^{-1} Q_1\) has a double eigenvalue.
This leads to the expression
\begin{align}
  \label{eq:P3-discriminant}
  \Delta = \operatorname{disc}_{\lambda} \det(\lambda - Q_2^{-1} Q_1)
\end{align}
for the elliptic discriminant.
Moreover, a defining equation for the curve is given by \(y^2 = \det(x- Q_2^{-1} Q_1) = 0\).
This depends rationally on the kinematic variables contained in \(Q_1\) and \(Q_2\) and a Weierstrass form and the \(j\)-invariant can subsequently be computed by rational transformations.

It turns out that the elliptic curve obtained in this way has the \emph{same} \(j\)-invariant as those computed from twistor space in ref.~\cite{Vergu:2020uur} and from the parametric representation of ref.~\cite{Bourjaily:2017bsb}. As we do not need to combine distinct square roots in this representation, this is consistent with our observations in the previous section.

In the submission of this paper to the arXiv, we have attached the file \texttt{doublebox\textunderscore{}curve.txt} that contains an expression for the defining equation of the curve.
With minor modifications the file should be readable with most computer programs.

\subsection{Four-dimensional derivation of the Baikov form}
\label{sec:4d-baikov}

In this section we present a derivation of the Baikov form without using dimensional regularization.  This avoids having to take the potentially somewhat tricky limit \(d \to 4\).  Equivalently, one can obtain the cut integrand as a one-form and it is not necessary to take one extra residue as in sec.~\ref{sec:double-box-baikov}.

Consider the loop parametrized by \(k_2\) in the elliptic double-box.  This loop has denominators
\begin{gather}
    D_1 = k_2^2, \qquad
    D_2 = (p_1 + k_2)^2, \qquad
    D_3 = (p_{12} + k_2)^2, \qquad
    D_4 = (k_1 - k_2)^2.
\end{gather}
It has ``external'' momenta \(p_1\), \(p_2\), \(k_1 + p_{12}\) and \(k_1\).  The integral measure \(\id^d k_2\) decomposes into an integral \(\id^3 k_2^\parallel\) over the space spanned by the independent ``external'' momenta \(p_1\), \(p_2\) and \(k_1\), and an orthogonal integral \(\id^{d - 3} k_2^\perp\).  The dot products of \(k_2\) with the ``external'' momenta are
\begin{gather}
    k_2 \cdot p_1 = \frac 1 2 (D_2 - p_1^2 - D_1), \\
    k_2 \cdot p_2 = \frac 1 2 (D_3 - D_2 - p_{12}^2 + p_1^2), \\
    k_2 \cdot k_1 = -\frac 1 2 (D_4 - D_1 - k_1^2).
\end{gather}
Using identities from appendix~\ref{sec:4d-baikov-appendix}, it follows that
\begin{gather}
    \id^3 k_2^\parallel = \frac {\id (k_2 \cdot p_1) \id (k_2 \cdot p_2) \id (k_2 \cdot k_1)}{\det G(p_1, p_2, k_1)^{\frac 1 2}} =
    -\frac 1 8 \frac {\id D_2 \id D_3 \id D_4 + \id D_1 (\cdots) }{\det G(p_1, p_2, k_1)^{\frac 1 2}}, \\
    \id^{d - 3} k_2^\perp = \frac 1 2 \Omega_{d - 3} \left(\frac {\det G(k_2, p_1, p_2, k_1)}{\det G(p_1, p_2, k_1)}\right)^{\frac {d - 5}2} \id D_1.
\end{gather}
Of course, we do not need to keep the dimension \(d\) arbitrary and we can set \(d = 4\) here.  In that case we have \(\Omega_1 = 2\).

When computing the full \(\id^4 k_2\) measure the extra terms in \(\id^3 k_2^\parallel\) proportional to \(\id D_1\) drop out:
\begin{equation}
    \id^4 k_2 = -2 \frac 1 {16} \id D_1 \id D_2 \id D_3 \id D_4 \bigl(\det G(p_1, p_2, k_1)\bigr)^{-\frac 1 2} \left(\frac {\det G(k_2, p_1, p_2, k_1)}{\det G(p_1, p_2, k_1)}\right)^{-\frac 1 2}.
\end{equation}
Note that we have not canceled the factor \(\det G(p_1, p_2, k_1)\) since we do not allow ourselves to combine square roots.  Note also that we have some Gram determinants whose entries contain \(k_1 \cdot p_1\), \(k_1 \cdot p_2\) and \(k_1^2\).  We need to keep these dot products in mind when analyzing the \(k_1\) integral, to which we turn next. 

For the \(k_1\) integral we have new denominators
\begin{gather}
    D_5 = (k_1 + p_{123})^2, \qquad
    D_6 = (k_1 + p_{1234})^2, \qquad
    D_7 = (k_1 - p_6)^2,
\end{gather}
while in the Jacobian of the \(\id^d k_2\) integral we have \(k_1^2\), \(k_1 \cdot p_1\) and \(k_1 \cdot p_2\).  We introduce two new Lorentz-invariant quantities \(D_8 = k_1^2\) and \(D_9 = (k_1 + p_{12})^2\).

However, not all quantities \(D_5, \dotsc, D_9\) can be independent; there are five such quantities and only four components for the vector \(k_1\).  The relation connecting these quantities can be obtained by computing the Gram determinant \(\det G(k_1, p_{12}, p_{123}, p_{1234}, p_{12345}) = 0\).  Equivalently, we can antisymmetrize in five different vectors to obtain
\begin{multline}
    k_1^\mu \epsilon(p_{12}, p_{123}, p_{1234}, p_{12345}) -
    p_{12}^\mu \epsilon(k_1, p_{123}, p_{1234}, p_{12345}) +
    p_{123}^\mu \epsilon(k_1, p_{12}, p_{1234}, p_{12345}) -\\
    p_{1234}^\mu \epsilon(k_1, p_{12}, p_{123}, p_{12345}) +
    p_{12345}^\mu \epsilon(k_1, p_{12}, p_{123}, p_{1234}) = 0.
\end{multline}

When decomposed over the basis \(p_{12}\), \(p_{123}\), \(p_{1234}\) and \(p_{12345}\), \(k_1\) has components \(k_1 \cdot p_{12}\), etc., with a metric given by the inverse of the Gram matrix \(G(p_{12}, p_{123}, p_{1234}, p_{12345})\).  The scalar products \(k_1^2\), \(k_1 \cdot p_1\) and \(k_1 \cdot p_2\) can be computed from this decomposition.  In particular, this implies that we can compute \(D_8 = k_1^2\) in terms of the other \(D_i\) (since here there are no transversal components there is no need to introduce \(D_8\) at all).  Let us compute the measure \(\id^4 k_1\) in terms of \(D_9\), \(D_5\), \(D_6\) and \(D_7\).  Using eq.~\eqref{eq:v_in_terms_of_D}, we find
\begin{equation}
    \id^4 k_1 = \frac 1 {2^4} \frac {(\det M_0)^{-\frac 1 2} \det M_0}{\bigl(\det M_1 \det M_0\bigr)^{\frac 1 2}} \id D_9 \id D_5 \id D_6 \id D_7,
\end{equation}
where
\begin{gather}
    M_0 = G(p_{12}, p_{123}, p_{1234}, p_{12345}), \\
    M_1 = \begin{pmatrix}
      D_9 & \frac 1 2 (D_9 + D_5 - p_3^2) & \frac 1 2 (D_9 + D_6 - p_{34}^2) & \frac 1 2 (D_9 + D_7 - p_{345}^2) \\
          & D_5 & \frac 1 2 (D_5 + D_6 - p_4^2) & \frac 1 2 (D_5 + D_7 - p_{45}^2) \\
          & & D_6 & \frac 1 2 (D_6 + D_7 - p_5^2) \\
          & & & D_7
    \end{pmatrix}.
\end{gather}
Here we have written only some of the matrix entries, the others can be determined from these by symmetry.

When taking the cuts we need to set \(D_1\) through \(D_7\) to zero, and thus we only need the expression for \(\det M_1\) when \(D_1 = \cdots = D_7 = 0\).  Then \(\det M_1\) is a quadratic polynomial in \(D_9\).  Taking the squares of the Jacobians obtained in this section we obtain a genus-one curve as an intersection of two quadrics.  This curve has the same \(j\)-invariant as the one obtained by considering the curve embedded in momentum twistor space as described in ref.~\cite{Vergu:2020uur}.

\section{Conclusions}
\label{sec:conclusions}

We have shown that the maximal cut and the Feynman parametrization of the two-loop sunrise integral do not necessarily correspond to different elliptic curves. The observation of different curves for these two objects in the literature was an artifact due to combining two square roots, and a more careful treatment shows the same curve for both the Feynman-parametric and Baikov representation, reinforced by the observation of the same curve in a light-cone parametrization of the maximal cut. We have shown that similarly the Baikov and twistor representations of the elliptic double-box also describe the same elliptic curve.

In some ways, the appearance of the same curve in different representations of these integrals should not be surprising. If one thinks of the maximal cut as a variety in loop momentum space, that variety should already define an elliptic curve. Whether we parametrize it with Baikov, light-cone, or twistor coordinates, we are performing changes of variables which should preserve invariant features of the geometry, such as the $j$-invariant. From this perspective, the surprise is actually that this curve is preserved in Feynman parameters. Feynman parameters do not correspond straightforwardly to a change of variables from the initial loop momenta, so the fact that they apparently preserve the geometry deserves further explanation.

One of the implications of our work is that analytic continuation of the Baikov representation away from from the Baikov integration domain has to be done with some care.  Inside this domain the Jacobians involved in changing coordinates are positive and one can pick the positive solution of any square roots that appear.  However, while this is possible for Euclidean kinematics, there is no canonical choice of square roots outside this region.

In ref.~\cite{Bourjaily:2020hjv}, an extension of the notion of leading singularity was put forward which applies to integrals containing genus-one curves as well.  The construction in that reference implicitly assumes a fixed geometry for the genus-one curve.  If there were a genuine ambiguity in the underlying genus-one curve it is not clear how one should modify their construction.  Fortunately, the results of this paper imply that such a modification may not be necessary.

In previous investigations of the elliptic double-box, conformal symmetry served as an important constraint that allowed for particularly clean representations. The Baikov representation is by its nature not conformal, as it uses momentum invariants as variables. It would be interesting to find a variant of Baikov that preserves conformal symmetry, to make better use of this kind of representation in the context of, e.g., $\mathcal{N}=4$ super Yang-Mills.

Finally, there is a broader concern raised by the observations of refs.~\cite{Adams:2017ejb,Bogner:2019lfa} that we do not fully address. While we do find the same curve for both the cut and Feynman parametrization of the sunrise integral, this by no means shows that isogenies are never relevant to the elliptic integrals that occur in physics. In particular, while our work suggests that each elliptic Feynman integral has a preferred curve, it may be that there exist distinct diagrams whose corresponding curves are isogenous. If such an example were to be found, it would suggest the need for a formalism that relates not merely iterated integrals on the same elliptic curve, but iterated integrals on isogenous curves as well.

\subsection*{Acknowledgements}

We thank Stefan Weinzierl for helpful discussions.
This work was supported in part by the Danish Independent Research Fund under grant number DFF-4002-00037 (MV), the Danish National Research Foundation (DNRF91), the research grant 00015369 from Villum Fonden, a Starting Grant (No.\ 757978) from the European Research Council (MV, MvH, CV) and the European Union’s Horizon 2020 research and innovation program under grant agreement No. 793151 (MvH). This project has received funding from the European Union’s Horizon 2020 research and innovation program under the Marie Sk{\l}odowska-Curie grant agreement No. 847523 ‘INTERACTIONS’ (HF). The work of HF has been partially supported by a Carlsberg Foundation Reintegration Fellowship.

\appendix

\section{Baikov representations with derivations}
\label{sec:baikov-dervations-appendix}

In this appendix we carefully derive the Baikov representation in its loop-by-loop and its standard forms. This derivation mostly follows ref.~\cite{Grozin:2011mt} and the loop-by-loop part additionally ref.~\cite{Frellesvig:2017aai}.

\subsection{The one-loop case}
\label{sec:baikov-derivations-oneloop}

As both the loop-by-loop and standard Baikov representations build off of the Baikov representation at one loop, we will start by reviewing the situation there. Writing a generic one-loop integral,

\begin{align}
I &= \int \frac{\id^d k}{i \pi^{d/2}}\frac{N(k)}{P_1(k)^{a_1} \cdots P_{\text{P}}(k)^{a_{\text{P}}}}
\end{align}
we then split the integral up in parts parallel and perpendicular to the space spanned by the $E$ independent external momenta:
\begin{align}
\id^d k &= \id^E k_{\parallel} \id^{d-E} k_{\perp} \\
&= \id^E k_{\parallel} \, |k_{\perp}|^{d-E-1} \, \id |k_{\perp}| \, \id^{d-E-1} \Omega.
\label{eq:full-measure}
\end{align}
Using
\begin{align}
\int \id^{n-1} \Omega = \Omega_n = \frac{2 \pi^{n/2}}{\Gamma(n/2)}
\end{align}
we get
\begin{align}
I &= \frac{2}{\Gamma((d-E)/2) \, i \pi^{E/2}} \int \frac{N (k) \; \id^E k_{\parallel} \, |k_{\perp}|^{d-E-1} \, \id |k_{\perp}|}{P_1(k)^{a_1} \cdots P_{\text{P}}(k)^{a_{\text{P}}}}.
\label{eq:Bfirsstep}
\end{align}

We may write the parallel component as
\begin{align}
  k_\parallel = \sum_{i=1}^E z_i p_i,
\end{align}
which implies that
\begin{align}
  k_\parallel \cdot p_j = k \cdot p_j = \sum_{i=1}^E z_i p_i \cdot p_j.
\end{align}
We introduce the Gram matrix \(G\) with entries \(G_{i j} = p_i \cdot p_j\). This allows us to write,
\begin{align}
  z_i = \sum_{j = 1}^E G^{-1}_{i j} (k \cdot p_j).
\end{align}

We further have that
\begin{align}
  k_\parallel^2 &= \sum_{i, j = 1}^E z_i z_j G_{i j} =
  \sum_{i, j = 1}^E (k \cdot p_i) (G^{-1})_{i j} (k \cdot p_j).
\end{align}

We may pick a basis in which the quantities
\begin{align}
\varsigma_i &:= k \cdot p_i \,.
\end{align}
are the components of the vector \(k_\parallel\).  In that case, the metric is nontrivial and is given by the inverse of the Gram matrix.  The integration measure is then
\begin{align}
  \id^{E} k_\parallel = (\det G^{-1})^{\frac 1 2} \prod_{i = 1}^E \id \varsigma_i.
 \label{eq:Bdkpara}
\end{align}
The orthogonal part has norm \(k_\perp^2 = k^2 - k_\parallel^2\).  Including the expression for \(k_\parallel^2\) we have
\begin{align}
    k_\perp^2 &= k^2 - \sum_{i, j = 1}^E (k \cdot p_i) (G^{-1})_{i j} (k \cdot p_j).
\label{eq:Bkperpsq}
\end{align}

Let us form the \((E + 1) \times (E + 1)\) Gram matrix,
\begin{align}
    \hat{G} = \begin{pmatrix}
    k^2 & k \cdot p_i \\
    k \cdot p_j & G_{j i}
    \end{pmatrix}.
\end{align}
Using the expression for the determinant of a matrix written in terms of blocks, we have that
\begin{align}
    \det \hat{G} = \left[k^2 - \sum_{i, j = 1}^E (k \cdot p_i) (G^{-1})_{i j} (k \cdot p_j)\right] \det G.
    \label{eq:kperp-expansion}
\end{align}
Hence, \((k_\perp)^2 = \frac {\det \hat G}{\det G}\).

Using the expression of \(k_\perp^2\) from eq.~\eqref{eq:kperp-expansion}, we find that \(\lvert k_\perp\rvert \id \lvert k_\perp\rvert = \lvert k\rvert \id \lvert k\rvert + \dotsc\), where the missing terms contain components \(\id \varsigma_i\) which vanish when wedged into \(\id^E k_\parallel\).
This means that we get the relation
\begin{align}
    \id |k_{\perp}| \id^E k_{\parallel} = \tfrac{1}{2} |k_{\perp}|^{-1} \id \varsigma_0 \, \id^E k_{\parallel}
    \label{eq:Bdkperp}
\end{align}
where we have used the notation $\varsigma_0 = k^2$.

Inserting eqs.~(\ref{eq:Bdkperp}, \ref{eq:Bkperpsq}, \ref{eq:Bdkpara}) into eq.~\eqref{eq:Bfirsstep}
we get
\begin{align}
I &= \frac{\mathcal{G}^{(E-d+1)/2}}{\Gamma((d-E)/2) \, i \pi^{E/2}} \int \frac{N (\varsigma) \; \mathcal{B}(\varsigma)^{(d-E-2)/2} \; \id^{E+1} \varsigma }{P_1(\varsigma)^{a_1} \cdots P_{\text{P}}(\varsigma)^{a_{\text{P}}}},
\end{align}
where we have defined
\begin{align}
\mathcal{B} := \det \hat{G} = \det G(k,p_1,\ldots,p_E)\;, \qquad \mathcal{G} := \det G = \det G(p_1,\ldots,p_E)\;,
\end{align}
with $G$ denoting the Gram matrix.

Now the only step left is to change to the Baikov variables $x_i$, which equal the propagators. If there are too few propagators ($\text{P} < E+1$) one will need to introduce additional variables, but this is mostly relevant at higher loops. The Jacobian $\mathcal{J}$ for the change of variables will depend on the exact expressions used for the propagators, but for most conventions it equals,
\begin{align}
\mathcal{J} &= \pm \, 2^{-E}.
\end{align}
Thus the final result for a one-loop Baikov representation is
\begin{align}
I &= \frac{\mathcal{J} \, \mathcal{G}^{(E-d+1)/2}}{\Gamma((d-E)/2) \, i \pi^{E/2}} \int \frac{N (x) \; \mathcal{B}(x)^{(d-E-2)/2} \; \id^{E+1} x }{x_1^{a_1} \cdots x_{\text{P}}^{a_{\text{P}}}}.
\label{eq:oneloop}
\end{align}

\subsection{Multi-loop, the loop-by-loop approach}
\label{sec:baikov-derivations-loop-by-loop}

With this representation in hand, we now want to apply it to multi-loop cases. A multi-loop Feynman integral is given by
\begin{align}
I &= \int \frac{\id^d k_1}{i \pi^{d/2}} \cdots \frac{\id^d k_L}{i \pi^{d/2}} \frac{N(\{k\})}{P_1(\{k\})^{a_1} \cdots P_{\text{P}}(\{k\})^{a_{\text{P}}}}
\end{align}
Our strategy will be to go through the steps from the previous section one loop at a time, starting with loop number $L$ and then going down towards $1$. We call $E_l$ the number of momenta external to loop number $l$. This may include the loop momenta of lower-numbered loops. We will denote with $\mathcal{G}_l$ the Gram-matrix of the momenta external to loop $l$, while $\mathcal{B}_l$ is the same but with the loop-momentum $k_l$ included.
If we follow the steps of the previous section with this notation, we arrive at the correspondence
\begin{align}
\frac{\id^d k_l}{i \pi^{d/2}} \; \rightarrow \; \frac{\mathcal{G}_l^{(E_l-d+1)/2} \, \mathcal{B}_l(\varsigma_l)^{(d-E_l-2)/2} }{\Gamma((d-E_l)/2) \, i \pi^{E_l/2}} \id^{E_l+1} \varsigma_l
\end{align}
where $\varsigma_l$ corresponds to the set of dot-products between $k_l$ and itself along with the momenta external to the $l$th loop. Putting this together for each loop gives
\begin{align}
I &= \frac{(-i)^L \, \pi^{-(\sum_i \! E_i)/2}}{\prod_l^L \Gamma((d-E_l)/2)} \int \frac{N (\varsigma) \left( \prod_l^L \mathcal{G}_l^{(E_l-d+1)/2} \, \mathcal{B}_l^{(d-E_l-2)/2} \right) \id^{(\sum_i \! E_i)+L} \varsigma }{P_1(\varsigma)^{a_1} \cdots P_{\text{P}}(\varsigma)^{a_{\text{P}}}}
\end{align}
and changing to the Baikov variables gives the final expression for the loop-by-loop Baikov representation:
\begin{align}
I &= \frac{\mathcal{J} \; (-i)^L \, \pi^{-(\sum_i \! E_i)/2}}{\prod_l^L \Gamma((d-E_l)/2)} \int \frac{N (x) \left( \prod_l^L \mathcal{G}_l^{(E_l-d+1)/2} \, \mathcal{B}_l^{(d-E_l-2)/2} \right) \id^{(\sum_i \! E_i)+L} x }{x_1^{a_1} \cdots x_{\text{P}}^{a_{\text{P}}}}
\label{eq:loop-by-loop}
\end{align}
where the Jacobian for the final variable change still depends on the specific expressions used for the propagators, but is usually given as
\begin{align}
\mathcal{J} &= \pm \, 2^{-(\sum_i \! E_i)}.
\end{align}
The expression of eq.~\eqref{eq:loop-by-loop} may also be found in ref.~\cite{Frellesvig:2021vem}.

\subsection{Multi-loop, the standard approach}
\label{sec:baikov-derivations-standard}

The standard approach to multi-loop Baikov parametrization can be thought of as a version of the loop-by-loop approach, but with the assumption that all loops depend on all lower loop-momenta and all external momenta. This means
\begin{align}
E_l &= E + l - 1
\end{align}
If this is the case then $\mathcal{G}_l = \mathcal{B}_{l-1} $ since their definitions will be the same.
We also have that the power of $\mathcal{G}_l$ which appears in the expression, $(E_l-d+1)/2$, is equal to minus the power with which $\mathcal{B}_{l-1}$ appears, making the two contributions cancel. This will happen pairwise for each loop, leaving only $\mathcal{B}_{L}$ and $\mathcal{G}_{1}$. Renaming these to $\mathcal{B}$ and $\mathcal{G}$ means we have
\begin{align}
\mathcal{B} = \det G( p_1, \ldots, p_E, k_1, \ldots, k_L) \qquad \text{and} \qquad \mathcal{G} = \det G( p_1, \ldots, p_E ).
\end{align}
Then eq.~\eqref{eq:loop-by-loop} becomes
\begin{align}
I &= \frac{\mathcal{J} \; (-i)^L \, \pi^{L-n} \, \mathcal{G}^{(E-d+1)/2}}{\prod_{l=1}^L \Gamma((d{+}1{-}E{-}l)/2)} \int \frac{N(x) \; \mathcal{B}^{(d{-}E{-}L{-}1)/2} \, \id^{n} x }{x_1^{a_1} \cdots x_{\text{P}}^{a_{\text{P}}}}
\label{eq:standard}
\end{align}
where we have used and defined
\begin{align}
n &\equiv L+\sum_i E_i \,=\, EL + L(L+1)/2
\end{align}
and where we (usually) have $\mathcal{J} = \pm 2^{L-n}$. We see that for $L=1$ eq.~\eqref{eq:standard} reduces nicely to eq.~\eqref{eq:oneloop}.

\subsection{The elliptic double-box}
\label{sec:baikov-derivations-doublebox}

Let us look at the example of the elliptic double-box shown in fig.~\ref{fig:double-box}.
We have the propagating momenta
\begin{align}
q_1 &= k_2, & q_2 &= k_2 + p_1, & q_3 &= k_2 + p_{12}, \nonumber \\
q_4 &= k_1 + p_{123}, & q_5 &= k_1 + p_{1234}, & q_6 &= k_1 + p_{12345}, \\
q_7 &= k_1 - k_2, & q_8 &= k_1, & q_9 &= k_1 + p_{12}. \nonumber
\end{align}
The last two $q_8$ and $q_9$ do not actually appear in the diagram, but they are needed to express all scalar products in terms of the Baikov variables.

We have $E_2=3$, counting the three momenta $k_1, p_1, p_2$ that are external to the $k_2$-loop, while $E_1=4$ since this is the maximum number of independent momenta in four space-time dimensions.
The four Gram determinants appearing are
\begin{align}
\mathcal{B}_2 &= \det G(k_2,k_1,p_1,p_2), & \mathcal{G}_2 &= \det G(k_1,p_1,p_2), \nonumber \\
\mathcal{B}_1 &= \det G(k_1,p_3,p_4,p_5,p_6), & \mathcal{G}_1 &= \det G(p_3,p_4,p_5,p_6).
\end{align}
We have $\mathcal{J} = \pm 2^{-7} \Bigl(\frac {\det G(p_1, p_2, p_3, p_4)}{\det G(p_3, p_4, p_5, p_6)}\Bigr)^{\frac{1}{2}}$. Putting this together in eq.~\eqref{eq:loop-by-loop} we obtain the expression
\begin{align}
I &= \frac{\mathcal{J} \, \pi^{-7/2} \, \mathcal{G}_1^{(5-d)/2}}{\Gamma((d{-}3)/2) \, \Gamma((d{-}4)/2)} \int \frac{N \! (x) \, \mathcal{G}_2^{(4-d)/2} \, \mathcal{B}_2^{(d-5)/2} \, \mathcal{B}_1^{(d-6)/2} \id^{9} x }{x_1^{a_1} \cdots x_{7}^{a_{7}}}.
\end{align}

\subsection{Derivation of a four-dimensional Baikov representation}
\label{sec:4d-baikov-appendix}

In this section we consider the case when there is no orthogonal component \(k_\perp = 0\), which will be needed for our derivation of a Baikov representation in four dimensions.  We also introduce the vectors \(v_i\) which are defined from the denominators \(D_i = (k - v_i)^2\), corresponding to massless propagators.  We take all of these vectors to be nonvanishing.  In other words, we will use as new coordinates the quantities \(D_i\), but \(k^2\) will not be one of these coordinates.

Then, we have
\begin{equation}
    \id^d k = (\det G^{-1})^{\frac 1 2} \prod_{i = 1}^d \id (k \cdot v_i).
\end{equation}
We want to express this in terms of \(D_i = (k - v_i)^2\) instead of \(k \cdot v_i\).  We have
\begin{align}
    \prod_{i = 1}^d \id D_i =
    \prod_{i = 1}^d 2 (k - v_i) \cdot \id v &=
    2^d \prod_{i = 1}^d (k \cdot \id k - \id (k \cdot v_i)) \nonumber \\
    &= (-2)^d \left[\prod_{i = 1}^d \id (k \cdot v_i) -
    \sum_{j = 1}^d (-1)^{j - 1} (k \cdot \id k) \prod_{i \neq j} \id (k \cdot v_i)\right].
\end{align}
Plugging in \(k \cdot \id k = \sum_{k, l} (k \cdot v_k) (G^{-1})_{k l} \id (k \cdot v_l)\), we obtain
\begin{equation}
    \prod_{i = 1}^d \id D_i = (-2)^d \left[1 - \sum_{j, k=1}^d (k \cdot v_k) (G^{-1})_{k j}\right] \prod_{i = 1}^d \id (k \cdot v_i).
\end{equation}

Let us rewrite the Jacobian in a simpler way
\begin{equation}
    \left[1 - \sum_{j, k = 1}^d (k \cdot v_k) (G^{-1})_{k j}\right] \det G =
    \det \begin{pmatrix}
    1 & k \cdot v_j \\
    1 & G_{j i}
    \end{pmatrix} =
    \det_{i j} ((v_i - k) \cdot v_j).
\end{equation}
To compute this last determinant, consider the decomposition of \(k - v_i\) and \(k - v_j\) on the basis of vectors \(v_k\).  Upon taking the dot product in this basis we obtain
\begin{equation}
    (k - v_i) \cdot (k - v_j) = \sum_{k, l = 1}^d ((k - v_i) \cdot v_k) (G^{-1})_{k l} ((k - v_j) \cdot v_l),
\end{equation}
whence
\begin{equation}
    \det_{i j} ((k - v_i) \cdot (k - v_j)) =
    \bigl(\det_{i k} ((k - v_i) \cdot v_k)\bigr)^2 (\det G)^{-1}.
\end{equation}

Since
\begin{equation}
    (k - v_i) \cdot (k - v_j) = \frac 1 2 \left[(k - v_i)^2 + (k - v_j)^2 - (v_i - v_j)^2\right] = \frac 1 2 \left[D_i + D_j - (v_i - v_j)^2\right],
\end{equation}
the determinant \(\det_{i j} ((k - v_i) \cdot (k - v_j))\) can be written in terms of \(D\) variables and constants \((v_i - v_j)^2\).  This determinant is the Cayley-Menger determinant which arises when computing the volume of a simplex in Euclidean space.

In the end, we find
\begin{equation}
    \label{eq:v_in_terms_of_D}
    \id^d k = \frac {(-1)^d}{2^d} \frac {(\det G)^{-\frac 1 2} \det G}{\bigl(\det_{i j} ((k - v_i) \cdot (k - v_j)) \det G\bigr)^{\frac 1 2}} \prod_{i = 1}^d \id D_i.
\end{equation}

\bibliographystyle{physics}
\bibliography{references}

\end{document}